\documentclass[3p,times,twocolumn]{elsarticle}
\biboptions{comma,sort&compress}

\usepackage[english]{babel}
\usepackage{graphicx,dcolumn,bm,amssymb,amsmath,amsfonts,amsthm,latexsym,widetext}
\usepackage{float,subfig,slashed,hyperref}
\hypersetup{
  linktocpage  = true,
  colorlinks   = true,
  urlcolor     = black,
  linkcolor    = black,
  citecolor    = black
}
\usepackage{ecrc}
\volume{00}
\firstpage{1}
\journalname{Nuclear and Particle Physics Proceedings}
\runauth{}
\jid{nppp}
\jnltitlelogo{Nuclear and Particle Physics Proceedings}

\begin{document}
\begin{frontmatter}

\title{The continuum threshold and the Polyakov loop: A comparison between two deconfinement order parameters}
	\cortext[cor0]{Talk given at 20th International Conference in Quantum Chromodynamics (QCD 17),  3-7 july 2017, Montpellier - FR}
\author[label1,label2]{J.P.~Carlomagno}
	\ead{carlomagno@fisica.unlp.edu.ar}
	\address[label1]{IFLP, CONICET $-$ Dpto.\ de F\'{\i}sica, Universidad Nacional de La Plata, C.C. 67, 1900 La Plata, Argentina}
	\address[label2]{CONICET, Rivadavia 1917, 1033 Buenos Aires, Argentina}
\author[label3,label4,label5]{M.~Loewe\fnref{fn1}}
	\fntext[fn1]{Speaker, Corresponding author.}
	\ead{mloewe@fis.puc.cl}
	\address[label3]{Instituto de F\'isica, Pontificia Universidad Cat\'olica de Chile, Casilla 306, San\-tia\-go 22, Chile}
	\address[label4]{Centre for Theoretical and Mathematical Physics and Department of Physics,
University of Cape Town, Rondebosch 7700, South Africa}
	\address[label5]{Centro Cient\'{\i}fico-Tecnol\'ogico de Valpara\'{\i}so - CCTVAL, Universidad T\'ecnica Federico Santa Mar\'{\i}a,  Casilla 110-V, Valpara\'{\i}so, Chile}   

\pagestyle{myheadings}
\markright{ }
\begin{abstract}
We compare two order parameters for the deconfinement transition, induced by thermal and density effects, commonly used in the literature, namely the thermal and density evolution of the continuum threshold $s_{0}$, within the frame of the QCD sum rules, and the trace of the Polyakov loop $\Phi$ in the framework of a nonlocal $SU(2)$ chiral quark model. We include in our discussion the evolution of the chiral quark condensate, the parameter that characterizes the chiral symmetry restoration. We found that essentially both order parameters, $s_{0}$ and $\Phi$, provide the same information for the deconfinement transition, both for the zero and finite chemical potential cases. At zero density, the critical temperatures in both cases coincide exactly and, in the case of finite baryonic chemical potential $\mu$, we find evidence for the appearance of a quarkyonic phase.
\end{abstract}

\end{frontmatter}


\section{Introduction}
\label{intro}

In QCD, when quarks are placed in a medium, the color charge is screened due to density and temperature effects~\cite{Fukushima:2010bq}. 
If the density and/or the temperature increases beyond a certain critical value, one expects that the interactions between quarks will not be able to confine them inside a hadron, so that they are free to travel longer distances and deconfine. 
This transition from a confined to a deconfined phase is usually referred to as the deconfinement phase transition.

A separate phase transition is the realization of chiral symmetry, moving from a Nambu-Goldstone phase into a Wigner-Weyl phase. 
Based, on lattice QCD evidence~\cite{Bazavov:2016uvm} one expects these two phase transitions to take place at approximately the same temperature at zero chemical potential. 
At finite density these two transitions can arise at different critical temperatures. 
The result will be a quarkyonic phase, where the chiral symmetry is restored but the quarks and gluons remain confined. 

It has been customary to study the behavior of  the trace of the Polyakov loop (PL) $\Phi(T, \mu)$ (order parameter for deconfinement phase transition) and quark anti-quark chiral condensate $\langle\bar{\psi} \psi\rangle(T, \mu)$ (chiral symmetry restoration), as function of temperature and chemical potential.

The goal of our discussion is to compare the Polyakov loop order parameter with a QCD deconfinement parameter~\cite{Bochkarev:1986es}, that corresponds to the squared energy threshold, $s_0(T,\mu)$, for the onset of perturbative QCD (PQCD) in hadronic spectral functions. 
For an actual general review see Ref.~\cite{Ayala:2016vnt}.
Around this energy, and at zero temperature, the resonance peaks in the spectrum dissapear  or become very broad, approaching then  the PQCD regime.
With increasing temperature approaching the critical temperature for deconfinement, the spectral function should then be described entirely by PQCD.

When both $T$ and $\mu$ are nonzero, lattice QCD simulations cannot be used, because of the sign problem in the fermionic determinant. 
Therefore, one need to resort either to mathematical constructions to overcome the above limitation, or to model calculations. 

The two deconfinement order parameters mentioned before: $\Phi(T, \mu)$ and $s_0(T, \mu)$ can be used to realize a phenomenological description of the deconfinement transition at finite temperature and density.

The natural framework to determine $s_0$ has been that of QCD sum rules. 
This framework is based on the operator product expansion (OPE) of current correlators at short distances, extended beyond perturbation theory, and on Cauchy's theorem in the complex $s$-plane. 
The latter is usually referred to as quark-hadron duality. 
Vacuum expectation values of quark and gluon field operators effectively parametrize the effects of confinement. 
An extension of this method to finite temperature was first outlined in~\cite{Bochkarev:1986es}. 
 
To analyze the role of the PL, we will concentrate on nonlocal Polyakov$-$Nambu$-$Jona-Lasinio (nlPNJL) models (see~\cite{Carlomagno:2013ona,Carlomagno:2016bpu} and references therein), in which quarks move in a background color field and interact through covariant nonlocal chirally symmetric four point couplings. 
These approaches,  offer a common framework to study both the chiral restoration and deconfinement transitions. 
In fact, the nonlocal character of the interactions arises naturally in the context of several successful approaches to low-energy quark dynamics, and leads to a momentum dependence in the quark propagator that can be made consistent~\cite{Noguera:2008cm} with lattice results.

The aim of the present work is to study the relation between both order parameters for the deconfinement transition at finite temperature and chemical potential, $\Phi$ and $s_0$, using the thermal finite energy sum rules (FESR) with inputs obtained from nlPNJL models. 


\section{Finite energy sum rules}
\label{fesr}

We begin by considering the (charged) axial-vector current correlator at $T=0$
\begin{align}
   \Pi_{\mu\nu}(q^2) &= i\int d^4x \,e^{iq\cdot x}\,
   \langle 0| T(A_\mu(x) A_\nu(0))|0 \rangle, \nonumber \\
   &= - g_{\mu\nu} \, \Pi_1(q^2) + q_\mu q_\nu \Pi_0(q^2) \;,
\label{correlator}
\end{align}
where $A_\mu(x) =\ :\bar{u}(x) \gamma_\mu \gamma_5 d(x):$ is the axial-vector current, $q_\mu = (\omega, \vec{q})$ is the four-momentum transfer, and the functions $\Pi_{0,1}(q^2)$ are free of kinematical singularities. 
Concentrating on the function $\Pi_0(q^2)$ and writing the OPE beyond perturbation theory in QCD , one of the two pillars of the sum rule method,  one has
\begin{equation}
\Pi_0(q^2)|_{\mbox{\tiny{QCD}}} = C_0 \, \hat{I} + \sum_{N=1} C_{2N} (q^2,\mu^2) \langle \hat{\mathcal{O}}_{2N} (\mu^2) \rangle \;, 
\label{OPE}
\end{equation}
where $\mu^2$ is a renormalization scale. 
The Wilson coefficients $C_N$ depend on the Lorentz indices and quantum numbers of the currents. 
Finally, the local gauge invariant operators ${\hat{\mathcal{O}}}_N$, are built from the quark and gluon fields in the QCD Lagrangian. 
The vacuum expectation values of those operators ($\hat{\mathcal{O}}_{2N} (\mu^2)$), dubbed as condensates, parametrize nonperturbative effects and have to be extracted from experimental data or model calculations. 

The second pillar of the QCD sum rules technique is Cauchy's theorem in the complex squared energy $s$-plane and this allows us to establish the following FESR. 
For details, we refer the reader to Ref.~\cite{Ayala:2016vnt} and to the original article Ref.~\cite{Carlomagno:2016bpu}

\begin{align}
(-)^{N-1} C_{2N} \langle {\mathcal{\hat{O}}}_{2N}\rangle = 4 \pi^2 \int_0^{s_0} ds\, s^{N-1} \,\frac{1}{\pi} {\mbox{Im}} \Pi_0(s)|_{\mbox{\tiny{HAD}}}
\nonumber \\
- \frac{s_0^N}{N} \left[1+{\mathcal{O}}(\alpha_s)\right] \;\; (N=1,2,\cdots) \;.
\label{FESR}
\end{align}

For $N=1$, the dimension $d=2$ term in the OPE does not involve any condensate, as it is not possible to construct a gauge invariant operator of such a dimension from the quark and gluon fields. 
There is no evidence for such a term (at $T=0$) from FESR analyses of experimental data on $e^+ e^-$ annihilation and $\tau$ decays into hadrons \cite{Dominguez:1999xa, Dominguez:2006ct}. 
At high temperatures, though, there seems to be evidence for some $d=2$ term \cite{Megias:2009ar}.
However, the analysis to be reported here is performed at lower values of $T$, so that we can safely ignore this contribution in the sequel. 

The dimension $d=4$ term, a renormalization group invariant quantity, is given by 
\begin{equation}
C_4 \langle \hat{\mathcal{O}}_{4}  \rangle = 
\frac{\pi}{6} \langle \alpha_s G^2\rangle + 2 \pi^2 (m_u + m_d) \langle\bar{q} q \rangle \ .
\label{C4}
\end{equation}

The extension of this program to finite temperature is fairly straightforward~\cite{Bochkarev:1986es, Dominguez:1994re, Ayala:2011vs}, with the Wilson coefficients in the OPE, Eq.~(\ref{OPE}), remaining independent of $T$ at leading order in $\alpha_s$, and the condensates developing a temperature dependence. 

In the static limit ($\vec{q} \rightarrow 0$), to leading order in PQCD, and for $T\neq 0$ and $\mu \neq 0$  the  function $\Pi_0(q^2)|_{\mbox{\tiny{QCD}}}$ in Eq.~(\ref{correlator}) becomes $\Pi_0(\omega^2, T, \mu)|_{\mbox{\tiny{QCD}}}$; to simplify the notation we shall omit the $T$ and $\mu$ dependence in the sequel. 
A calculation of the spectral function in perturbative QCD, at finite temperature and finite density gives
\begin{align}
   \frac{1}{\pi} {\mbox{Im}}\Pi_0(s)|_{\mbox{\tiny{PQCD}}}
   =
   \frac{1}{4\pi^2}\left[1-\tilde{n}_+\left(\frac{\sqrt{s}}{2}\right) 
   -\tilde{n}_-\left(\frac{\sqrt{s}}{2}\right)\right] \nonumber \\
   -\frac{2}{\pi^2} \;T^2 \;\delta (s)\; \left[
   {\mbox{Li}}_2(-e^{\mu/T})
   + {\mbox{Li}}_2(-e^{-\mu/T})\right] \ ,
\label{pertQCD}
\end{align}
where ${\mbox{Li}}_2(x)$ is the dilogarithm function, $s=\omega^2$, and $    \tilde{n}_\pm(x)=\ (e^{(x\mp \mu)/T}+1)^{-1} $ are the Fermi-Dirac thermal distributions for particles and antiparticles,
respectively. 

In the hadronic sector we assume pion-pole dominance of the hadronic spectral function, i.e. the continuum threshold $s_0$ to lie below the first radial excitation with mass $M_{\pi_1} \simeq 1 300\;{\mbox{MeV}}$. 

We have, then 
\begin{equation}
   \frac{1}{\pi}{\mbox{Im}}\Pi_0 (s)|_{\mbox{\tiny{HAD}}}
   = 2 \; f_\pi^2(T,\mu_B)\; \delta (s-m_\pi^2) \ ,
\label{HAD}
\end{equation}
where $f_\pi(T,\mu_B)$ is the pion decay constant at finite $T$ and $\mu$, with $f_\pi(0,0) =92.21 \pm 0.14 \;{\mbox{MeV}}$ \cite{Agashe:2014kda}. 
Notice we will not include in our spectral function the first part of $a_1$ resonance obtained from the $\tau$-decay data~\cite{Dominguez:2012bs}, since still there is no counterpart in the SU(2) nlPNJL model for the description of the hadronic vector resonance. 

Turning to the FESR, Eq.~(\ref{FESR}), with $N=1$ and no dimension $d=2$ condensate, and using Eqs.~(\ref{pertQCD}) and (\ref{HAD}) one finds
\begin{align}
\frac{}{}\int_0^{s_0(T,\mu)}  ds \, \left[1 - \tilde{n}_+\left(\frac{\sqrt{s}}{2}\right) 
   -\tilde{n}_-\left(\frac{\sqrt{s}}{2}\right)\right] = \nonumber\\
   8 \pi^2 f_\pi^2(T,\mu) +  8 T^2 \left[{\mbox{Li}}_2(-e^{\mu/T})
   + {\mbox{Li}}_2(-e^{-\mu/T})\right] \ .
\label{FESRTMU}
\end{align}
This is a transcendental equation determining $s_0(T,\mu)$ in terms of $f_\pi(T,\mu)$.   The next thermal FESR at zero chemical potential, for completeness, is given by~\cite{Dominguez:2012bs},
\begin{align}
- C_{4}\langle {\mathcal{\hat{O}}}_{4}\rangle(T) = 4 \pi^2 \int_0^{s_0(T)} ds\, s \frac{1}{\pi} {\mbox{Im}}\, \Pi_0(s)|_{\mbox{\tiny{HAD}}}
\nonumber \\
-  \int_0^{s_0(T)}ds \, s \left[1 - 2  n_F\left(\frac{\sqrt{s}}{2 T}\right)\right] ,
\label{FESRT2}
\end{align}
where $n_F(x)=1/(1+e^x)$ is the Fermi thermal function.


\section{Thermodynamics at finite density in the PNJL model}
\label{thermo}

We consider a nonlocal SU(2) chiral quark model that includes quark couplings to the color gauge fields~\cite{Carlomagno:2016bpu}.  
The quark-antiquark currents include nonlocal covariant form factors ${\cal G}(z)$ and ${\cal F}(z)$ characterizing the corresponding four-fermion interactions. 
The scalar-isoscalar current will generate a momentum dependent quark mass in the quark propagator, while the ``momentum'' current will be responsible for a momentum dependent quark wave function renormalization (WFR)~\cite{Noguera:2008cm,Contrera:2010kz,Pagura:2011rt}. 

To proceed, we perform a bosonization of the theory, introducing bosonic fields $\sigma_{1,2}(x)$ and $\pi_a(x)$, and integrating out the quark fields. 
Details of this procedure can be found e.g.\ in Ref.~\cite{Noguera:2008cm}. 

In order to analyze the properties of meson fields it is necessary to consider the quadratic fluctuations in the Euclidean action:
\begin{align}
\label{spiketa}
S_E^{\rm quad} &=& \dfrac{1}{2} \int \frac{d^4 p}{(2\pi)^4} \sum_{M}\  r_M\
G_M(p^2)\  \phi_M(p)\, \bar\phi_M(-p) \ ,
\end{align}
where meson fluctuations $\delta\sigma_a$, $\delta\pi_a$  have been translated to a charged basis $\phi_M$, being $M$ the scalar and pseudoscalar mesons ($\sigma,\pi^0$, $\pi^\pm$) plus the $\sigma_2$ field, and $G_M$ are the inverse dressed propagators.
The coefficient $r_M$ is 1 for charge eigenstates $M=\sigma_i,\pi^0$, and 2 for $M=\pi^+$. 
At finite temperature, the meson masses are obtained by solving $G_M (- m_M^2, 0) = 0$. 
The full expressions for the one-loop functions $G_M(q)$ can be found in Ref.~\cite{Noguera:2008cm,Carlomagno:2013ona}. 

Following a standard procedure, we can finally identify the corresponding pion weak decay constant
\begin{equation}
f_{\pi}=\frac{m\; Z^{-1/2}_\pi}{m_{\pi}^{2}}\; F_{0}(-m_{\pi}^{2})\ .
\label{fpi}%
\end{equation}
with
\begin{align}
F_{0}(p^{2})=8\, N_{c}\int\frac{d^{4}q}{(2\pi)^{4}}\ g(q)\;\frac
{Z(q^{+})Z(q^{-})}{D(q^{+})D(q^{-})}\times \nonumber \\
\left[q^{+}\cdot q^{-}+M(q^{+})M(q^{-})\right]
\label{fpi2}
\end{align}
where $q^{\pm}=q\pm p/2\,$ and $D(q)=q^{2}+M^{2}(q)$, with $M(p)$ and $Z(p)$ defined as
\begin{align}
M(p) =  Z(p) \left[m + \bar\sigma_1  g(p) \right] \ , \quad
Z(p) =  \left[ 1 - \bar\sigma_2  f(p) \right]^{-1}\ , \nonumber
\end{align}
here $g(p)$ and $f(p)$ are the Fourier transforms of the form factors ${\cal G}(z)$ and ${\cal F}(z)$.

We extend the bosonized effective action to finite temperature $T$ and chemical potential $\mu$ using the standard imaginary time formalism. 
Concerning the gauge fields, we assume that quarks move on a constant background field $\phi = A_4 = i A_0 = i g\,\delta_{\mu 0}\, G^\mu_a \lambda^a/2$, where $G^\mu_a$ are SU(3) color gauge fields. 
Then the traced Polyakov loop, which in the infinite quark mass limit can be taken as an order parameter of confinement, is given by $\Phi=\frac{1}{3} {\rm Tr}\, \exp( i \phi/T)$. 
We work in the so-called Polyakov gauge~\cite{Diakonov:2004kc}, where the matrix $\phi$ is given a diagonal representation $\phi = \phi_3 \lambda_3 + \phi_8 \lambda_8$. 

With the constraint of $\phi_3$ and $\phi_8$ being real~\cite{Dumitru:2005ng,Roessner:2006xn}, implies $\phi_8=0$, leaving only $\phi_3$ as an independent variable, and therefore $\Phi = [ 2 \cos(\phi_3/T) + 1 ]/3$.

Following the same prescriptions as in Refs.~\cite{GomezDumm:2001fz,GomezDumm:2004sr,Carlomagno:2013ona}, the real part of  $\Omega^{\rm MFA}$ at finite temperature $T$ and chemical potential $\mu$ is given by
\begin{equation}
\label{omegareg}
\Omega^{\rm MFA} \ = \ \Omega^{\rm reg} + \Omega^{\rm free} +
\mathcal{U}(\Phi,T) + \Omega_0 \ ,
\end{equation}
where
\begin{align}
\small
\Omega^{\rm reg} =  &-\ 4 T \sum_{c,n} 
\int_{\vec p} \log \left[ \frac{ 
(\rho_{n,\vec{p}}^c)^2 + M^2(\rho_{n,\vec{p}}^c)}{Z^2(\rho_{n, \vec{p}}^c)}\right]\ +  \nonumber \\ 
&\frac{\bar\sigma_1^2 + \kappa_p^2 \bar\sigma_2^2}{2\,G_S} \ , \nonumber \\
\Omega^{\rm free} = &-\  4 T \int_{\vec p} \sum_{c, s=\pm 1}\mbox{Re}\;
\log \left[ 1 + \exp\left(-\;\frac{\epsilon_p + i s \phi_c}{T}
\right)\right]
\ ,
\label{granp}
\end{align}
here $\bar\sigma_{1,2}$ are the mean field values of the scalar fields. We have also defined
\begin{equation}
\Big({\rho_{n,\vec{p}}^c} \Big)^2 =
\Big[ (2 n +1 )\pi  T + \phi_c - \imath \mu \Big]^2 + {\vec{p}}\ \! ^2 \ , 
\end{equation}
the sums over color indices run over $c=r,g,b$, with the color background fields components being $\phi_r = - \phi_g = \phi_3$, $\phi_b = 0$, and $\epsilon_p = \sqrt{\vec{p}^{\;2}+m^2}\;$. 

One possible Ansatz for the Polyakov loop potential $\mathcal{U}(\Phi,T)$ is that based on the logarithmic expression of the Haar measure associated with the SU(3) color group integration~\cite{Roessner:2006xn}.
Besides the logarithmic function, a widely used potential is that given by a polynomial function based on a Ginzburg-Landau Ansatz~\cite{Scavenius:2002ru,Ratti:2005jh}.
The corresponding expressions can be found in Ref.~\cite{Carlomagno:2016bpu}. 

Given the full form of the thermodynamical potential, the mean field values $\bar\sigma_{1,2}$ and $\phi_{3}$ can be obtained as solutions of the coupled set of gap equations
\begin{equation}
\frac{\partial \Omega^{\rm MFA}_{\rm reg}}
{\left(\partial\sigma_{1},\partial\sigma_{2}, \partial\phi_3\right)}\ = \ 0 \ .
\label{fullgeq}
\end{equation}

In order to fully specify the model under consideration, we proceed to fix the model parameters as well as the nonlocal form factors $g(q)$ and $f(q)$. We consider here Gaussian functions~\cite{Carlomagno:2016bpu} which guarantee a fast ultraviolet convergence of the loop integrals. 
The values of the five free parameters can be found in Ref.~\cite{Noguera:2008cm}.

Once the mean field values are obtained, the behavior of other relevant quantities as functions of the temperature and chemical potential can be determined. 
We concentrate, in particular, on the chiral quark condensate $\langle\bar{q}q\rangle = \partial\Omega^{\rm MFA}_{\rm reg}/\partial m$ and the traced Polyakov loop $\Phi$, which will be taken as order parameters for the chiral restoration and deconfinement transitions, respectively. 
The associated susceptibilities will be defined as $\chi_{\rm ch}  = \partial\,\langle\bar qq\rangle/\partial m$ and $\chi_{\rm PL} = d \Phi / d T$. 

In this work we define the deconfinement transition temperature, in the crossover region, with the peak of the Polyakov susceptibility $\chi_{\rm PL}$.
In the region where the deconfinement is a first order phase transition we use the same prescription as Ref.~\cite{Contrera:2010kz}, where the critical temperature is defined as the temperature where $\Phi=0.4$.


\section{Results}
\label{results}

We begin our analysis studying the finite energy sum rules at zero density. 
In this scenario,  when $\mu=0$, the Eq.~(\ref{FESRTMU}) becomes
\begin{align}
8  \pi^2 f^2_\pi(T) &=& \frac{4}{3}  \pi^2  T^2  + \int_0^{s_0(T)}ds \,\left[1 - 2\, n_F \left(\frac{\sqrt{s}}{2 T} \right) \right] \;, \label{FESRT1}
\end{align}
where the pion decay constant at finite temperature and/or chemical potential is calculated using  Eq.~(\ref{fpi}) and Eq.~(\ref{fpi2}) as
\begin{align}
F_{0}(p^{2})=8\, T \sum_{c,n} \int\frac{d^{3}\vec{q}}{(2\pi)^{4}}\ g({\rho_{n,\vec{q}}^c})\; 
\frac{Z({\rho_{n,\vec{q}}^c}^{+})Z({\rho_{n,\vec{q}}^c}^{-})}{D({\rho_{n,\vec{q}}^c}^{+})D({\rho_{n,\vec{q}}^c}^{-})}\ \times \nonumber \\
\left[  {\rho_{n,\vec{q}}^c}^{+}\cdot {\rho_{n,\vec{q}}^c}^{-}+M({\rho_{n,\vec{q}}^c}^{+}%
)M({\rho_{n,\vec{q}}^c}^{-})\right]
\label{fpi3}
\end{align}
where ${\rho_{n,\vec{q}}^c}^{\pm}={\rho_{n,\vec{q}}^c} \pm p/2\,$.

It is known that in local versions of the PNJL model, at zero chemical potential, the restoration of the chiral symmetry and the deconfinement transition take place at different temperatures (see e.g.~Refs.~\cite{Fu:2007xc,Costa:2008dp}), usually separated by approximate $20$~MeV. 

\begin{figure}[h]
\centering
\includegraphics[width=0.4\textwidth]{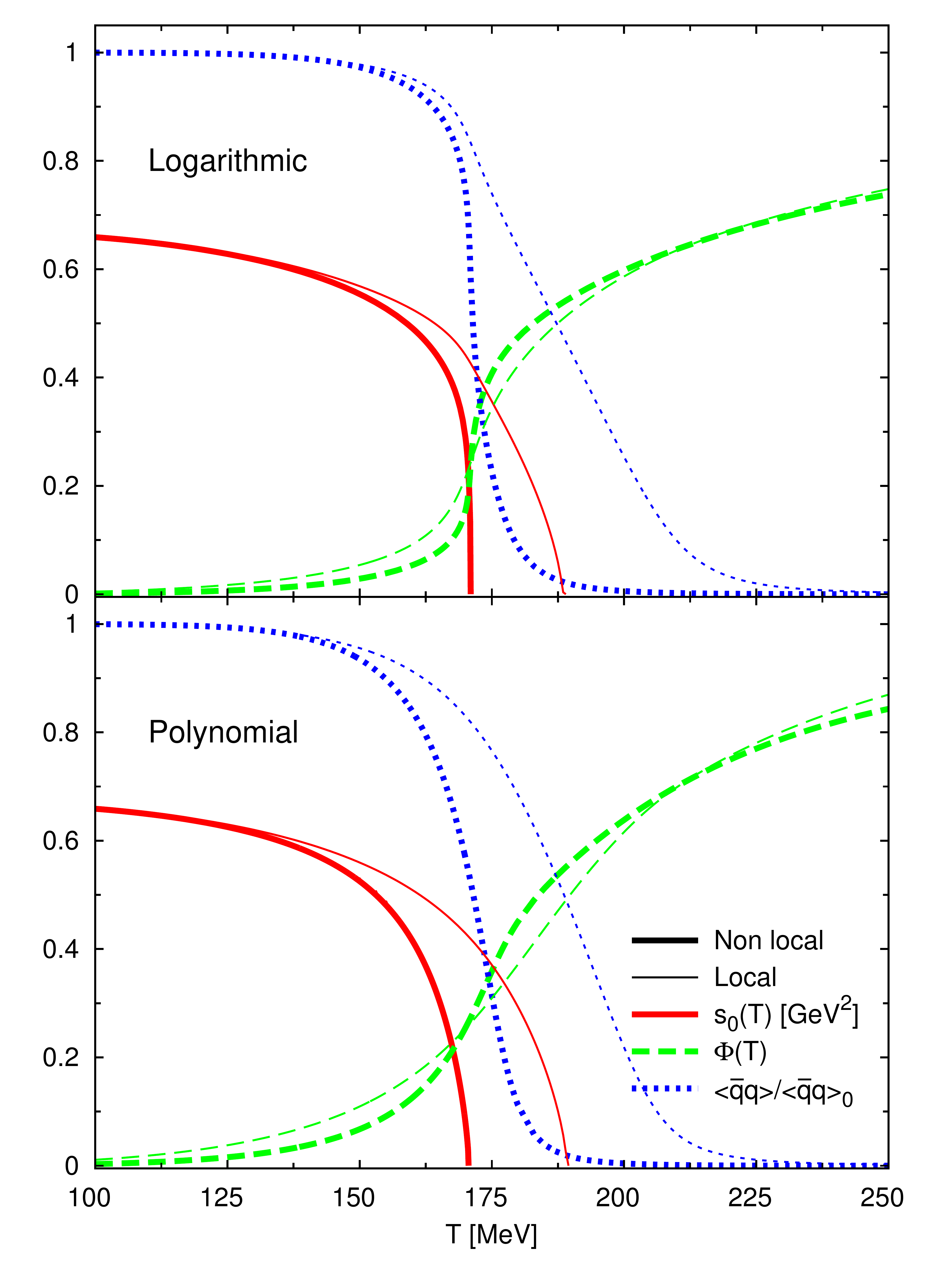}
\caption{\label{fig:lvsnl} Continuum threshold (red solid line), trace of the Polyakov loop (green dashed line) and the normalized quark condensate (blue dotted line) as a function of the temperature for nonlocal (thick line) and local PNJL model (thin line).}
\end{figure}

In Fig.~\ref{fig:lvsnl} we plot the continuum threshold, the trace of the PL and the normalized quark condensate for the nonlocal (local) PNJL model in thick (thin) line, for the logarithmic and polynomial effective potentials. 
As we expected from previous results, in the local version both transitions do not occur simultaneously. 
In this scenario, the PQCD threshold vanishes at a critical temperature, $T_c^{s_0}$, located between the chiral critical temperature $T_c^{\chi}$ and the PL deconfinement temperature $T_c^{\Phi}$ (obtained through the corresponding susceptibilities).

In the case of the nonlocal PNJL model, for both effective potentials, $s_0$ and $\Phi$ have a similar critical temperature for the deconfinement transition of approximate $T_c \sim 170$ MeV.
These temperatures are summarized in Table~\ref{TableI:tes_c}.
\begin{table}
\begin{tabular}{c c c c c}
\hline 
 & \multicolumn{2}{c}{Logarithmic} & \multicolumn{2}{c}{Polynomial} \\
\hline 
 & Non local & Local & Non local & Local \\
\hline
$T_c^{\chi}$ [MeV] & 171 & 205 & 176 & 201 \\
$T_c^{\Phi}$ [MeV] & 171 & 171 & 174 & 183 \\
$T_c^{s_0}$ [MeV]  & 171 & 189 & 170 & 190 \\
\hline
\end{tabular}
\caption{Chiral critical temperatures $T_c^{\chi}$ and deconfinement temperatures $T_c^{\Phi}$ and $T_c^{s_0}$.}
\label{TableI:tes_c}
\end{table}

From lattice QCD calculations, at zero chemical potential, the chiral symmetry restoration and the deconfinement transition take place at the same critical temperature. This behavior was verified in nlPNJL models~\cite{Contrera:2010kz,Carlomagno:2013ona,Carlomagno:2015hea} and also obtained by finite energy sum rules~\cite{Ayala:2011vs}. 
We will now identify the relation between $s_0(T,\mu)$ and $\Phi(T,\mu)$, extending our previous discussion.

\begin{figure}[h]
\centering
\includegraphics[width=0.4\textwidth]{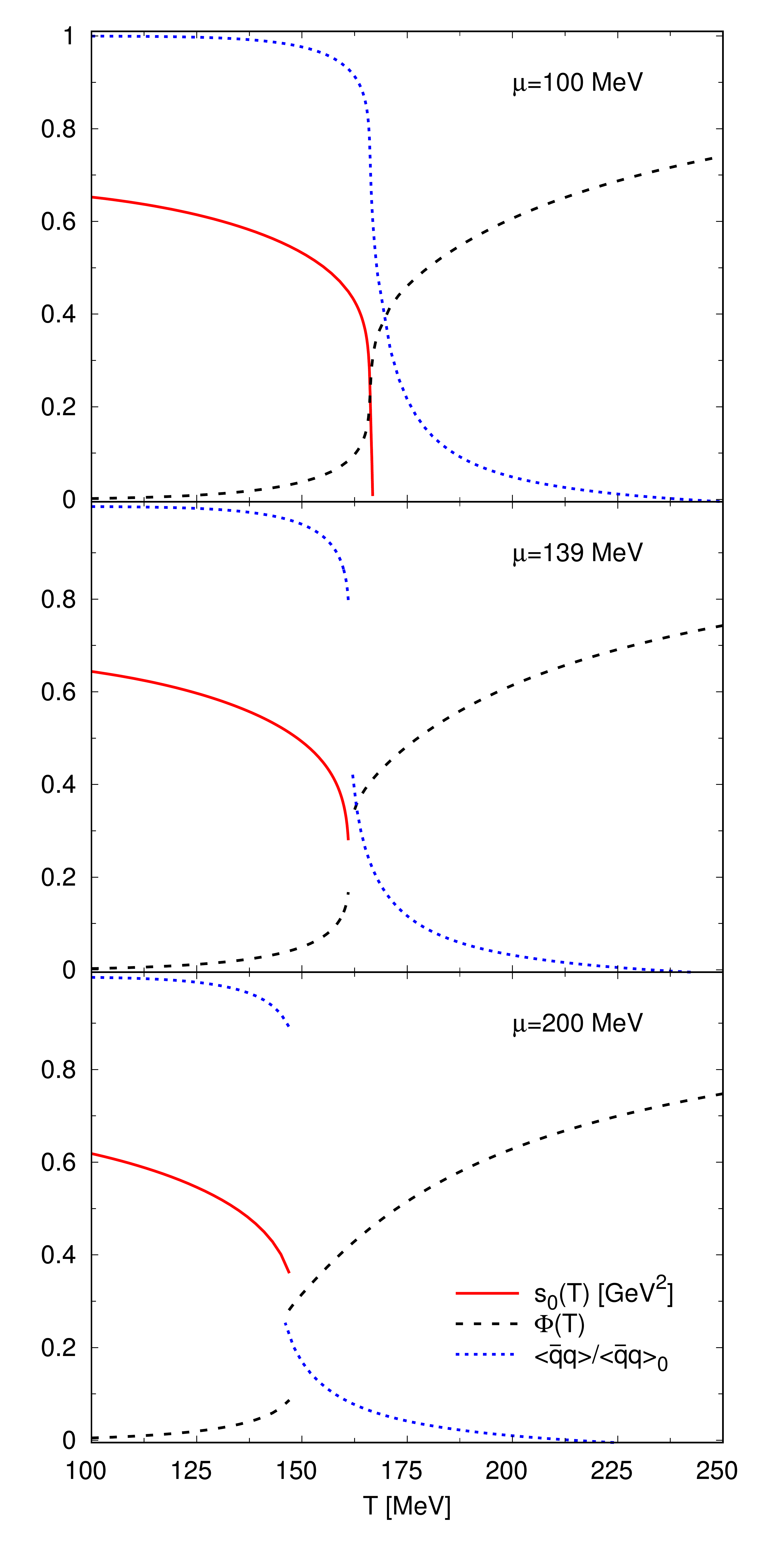}
\caption{\label{fig:munocero} Continuum threshold (solid red line), trace of the Polyakov loop (black dashed lined) and the normalized quark condensate (blue dotted line) as a function of the temperature for the logarithmic effective potential.}
\end{figure}

In Fig.~\ref{fig:munocero} we plot, for the logarithmic Polyakov effective potential, the normalized quark condensate $\langle\bar qq\rangle/\langle\bar qq\rangle_0$, the trace of the PL $\Phi$ and the continuum threshold $s_0$ as functions of the temperature for three different values of chemical potential. 
In the middle panel we choose $\mu=139$~MeV, which correspond to the critical end point chemical potential $\mu_{CEP}$. For values of $\mu$ smaller than $\mu_{CEP}$, the chiral restoration arises via a crossover transition. Beyond this critical density, a first order phase transition occurs. 
This value, together with the critical temperature $T_{CEP} = 161$~MeV determines the coordinates of the critical end point.

In the upper panel of Fig.~\ref{fig:munocero}, where $\mu=100$ MeV, we see that the chiral and deconfinement transitions are crossovers occurring at the same critical temperature. 
The peak of the Polyakov susceptibility and the point where the continuum threshold vanishes occur at approximate the same temperature $T_c \sim 166$ MeV.

When $\mu$ becomes equal or higher than $\mu=139$~MeV, the order parameter for the chiral symmetry restoration has a discontinuity signaling a first order phase transition. 
These gap in the quark condensate induces also a jump in the trace of the PL (see middle and lower panels in Fig.~\ref{fig:munocero}). The value of $\Phi$ at the discontinuity indicates that at this temperature the system remains confined but in a chiral symmetry restored state. This region is usually referred as the quarkyonic phase~\cite{McLerran:2007qj,Abuki:2008nm}.

We see in this way, that the Polyakov loop and the continuum threshold provide the same information. When the chiral symmetry is restored, $s_0$ and $\Phi$ show that we are still in a confined phase. This characterizes the occurrence of a quarkyonic phase.


\section{Summary and conclusions}
\label{finale}

In this article we compare the behavior of two order parameters for the deconfinement transition: the continuum threshold and the trace of the Polyakov loop.

To accomplish this analysis, we use finite energy sum rules for the axial-vector current correlator. 

On the other side, the Polyakov loop, is expected to vanish in the confined phase being different from zero in the deconfined phase.

By saturating the FESR with the pion pole in the spectral function, we used as an input the pion mass, the pion decay constant and the chiral quark condensate obtained from a nonlocal SU(2) Polyakov-NJL model with Gaussian form factors, establishing the connection between both approaches.

We determine, for the nlPNJL model, that the continuum threshold vanishes at the same temperature where the Polyakov susceptibility has its maximum value. 
In the case of the local PNJL, $s_0$ becomes zero between the critical temperature for the deconfinement transition, according to the Polyakov loop analysis, and the chiral restoration temperature. 

At finite chemical potential, we find that for both deconfinement parameters, beyond the critical end point chemical potential, the system remains in its confined phase even when the chiral symmetry is restored. This is an evidence for the appearance of a quarkyonic phase. 
 
We may conclude saying that our analysis gives strong support to the idea that both deconfinement parameters, in fact, provide the same kind of physical information.


\section*{Acknowledgements}

This work has been partially funded by CONICET (Argentina) under Grant No.\ PIP 449; by the National University of La Plata (Argentina), Project No.\ X718; by FONDECYT (Chile), under grants No. 1170107, 1150471 and 1150847; and by Proyecto Basal (Chile) FB 0821. 



\end{document}